\documentclass[aps,prd,12pt,preprint,tightenlines,superscriptaddress,
amsfonts,amssymb,amsmath,byrevtex,showpacs,nofootinbib]{revtex4}
\usepackage{graphics}
\usepackage{epsfig,subfigure}
\begin{document}
\newcommand{\dR}{\mathbb R}
\newcommand{\dC}{\mathbb C}
\newcommand{\dS}{\mathbb S}
\newcommand{\dZ}{\mathbb Z}
\newcommand{\dN}{\mathbb N}
\newcommand{\dQ}{\mathbb Q}
\newcommand{\id}{\mathbb I}
\newcommand{\ep}{\epsilon}
\newcommand{\dV}{\mathbb V}
\newcommand{\dH}{\mathbb H}
\newcommand{\dM}{\mathbb M}
\newcommand{\sing}{\mathcal{S}}
\newcommand{\up}{\uparrow}
\newcommand{\down}{\downarrow}
\renewcommand{\figurename}{Figure}

\newenvironment{narrow}[2]{%
\begin{list}{}{%
\setlength{\topsep}{0pt}%
\setlength{\leftmargin}{#1}%
\setlength{\rightmargin}{#2}%
\setlength{\listparindent}{\parindent}%
\setlength{\itemindent}{\parindent}%
\setlength{\parsep}{\parskip}}%
\item[]}{\end{list}}

\title{Probing the cosmological singularity with a particle}

\author{Przemys{\l}aw Ma{\l}kiewicz$^\dag$ and W{\l}odzimierz Piechocki$^\ddag$
\\ Department of Theoretical Physics\\So\l tan Institute for Nuclear Studies,
\\ Ho\.{z}a 69, 00-681 Warszawa, Poland;
\\ $^\dag$pmalk@fuw.edu.pl, $^\ddag$piech@fuw.edu.pl}

\date{\today}

\begin{abstract}

We examine the transition of a particle across the singularity of
the compactified Milne (CM) space. Quantization of the phase space
of a particle and testing the quantum stability of its dynamics
are consistent to one another.  One type of transition of a
quantum particle is described by a quantum state that is
continuous at the singularity. It indicates the existence of a
deterministic link between the propagation of a particle before
and after crossing the singularity. Regularization of the CM space
leads to the dynamics similar to the dynamics in the de Sitter
space. The CM space  is a promising model to describe  the
cosmological singularity deserving further investigation by making
use of strings and membranes.
\end{abstract}
 \maketitle

\section{Introduction}

Presently  available cosmological data indicate that known forms of
energy and matter comprise only $4\%$ of the makeup of the Universe.
The remaining $96\%$ is unknown, called `dark', but its existence is
needed to explain the evolution of the Universe
\cite{Spergel:2003cb,Bahcall:1999xn}.  The dark matter, DM,
contributes $22\%$ of the mean density. It is introduced to explain
the observed dynamics of galaxies and clusters of galaxies. The dark
energy, DE, comprises $74\%$ of the density and is responsible for
the observed accelerating expansion. These data mean that we know
almost nothing about the dominant components of the Universe!

Understanding the nature and the abundance of the DE and DM within
the standard  model of cosmology has  difficulties \cite{NGT,GDS}.
These difficulties have led many physicists to seek anthropic
explanations which, unfortunately, have little predictive power.
An alternative model has been proposed by Steinhardt and Turok
(ST) \cite{Steinhardt:2001vw,Steinhardt:2001st,Steinhardt:2004gk}.
It is based on the idea of a cyclic evolution, CE, of the
Universe. The ST model has been inspired by string/M  theories
\cite{Khoury:2001bz}. In its simplest version it assumes that the
spacetime can be modelled by the higher dimensional compactified
Milne, CM, space. The attraction of the ST model is that it
potentially provides a complete scenario of the evolution of the
universe, one in which the DE and DM play a key role in both the
past and the future. The ST model \textit{requires} DE for its
consistency, whereas in the standard model, DE is introduced in a
totally  \textit{ad hoc} manner. Demerits of the ST model  are
extensively discussed in \cite{Linde:2002ws}. Response to the
criticisms of \cite{Linde:2002ws} can be found in \cite{NGT}.

The mathematical structure and self-consistency of the ST model has
yet not been fully tested and understood. Such task presents a
serious mathematical challenge. It is the subject of our research
programme.

The CE model has in each of its cycles  a quantum phase including
the cosmological singularity, CS.  The CS plays key role because
it joins each two consecutive classical phases.  Understanding the
nature of the CS has  primary importance for the CE model.  Each
CS consists of contraction and expansion phases. A physically
correct model of the CS, within the framework of string/M theory,
should be able to describe  propagation of a  p-brane, i.e. an
elementary object like a particle, string and membrane, from the
pre-singularity to post-singularity epoch. This is the most
elementary, and fundamental, criterion that should be satisfied.
It presents a new criterion for testing the CE model. Hitherto,
most research has focussed on the evolution of scalar
perturbations through the CS.

Successful quantization of the dynamics of p-brane will mean that
the CM space is a promising candidate to model the evolution of
the Universe at the cosmological singularity. Thus, it could be
further used in advanced numerical calculations to explain  the
data of observational cosmology. Failure in quantization may mean
that the CS should be modelled by a spacetime more sophisticated
than the CM space.

Preliminary insight into the problem has already been achieved by
studying classical and quantum dynamics of a test particle in the
two-dimensional CM space \cite{Malkiewicz:2005ii}.  The present paper
is a continuation of \cite{Malkiewicz:2005ii} and it addresses the
two issues:  the Cauchy problem at the CS  and the stability problem
in the propagation of a particle across the CS. Both issues concern
the nature of the CS.

In Sec. II we define and make comparison of the two models of the
universe: the CM space and the regularized CM space. The classical
dynamics of a particle in both spaces is presented in Sec. III.  The
quantization of the phase space of a particle is carried out in Sec.
IV. In Sec. V we examine the stability problem of particle's dynamics
both at classical and quantum levels. We summarize our results,
conclude  and suggest next steps in Sec. VI.

\section{Spacetimes}

\subsection{The CM space}

For completeness, we recall  the definition of the CM space used in
\cite{Malkiewicz:2005ii}. It can be specified by the following
isometric embedding of the 2d CM space into the 3d Minkowski space
\begin{equation}\label{emb}
    y^0(t,\theta) = t\sqrt{1+r^2},~~~~y^1(t,\theta) =
    rt\sin(\theta/r),~~~~y^2(t,\theta) = rt\cos(\theta/r),
\end{equation}
where $ (t,\theta)\in \dR^1 \times \dS^1 $ and $ 0<r \in\dR^1 $ is a
constant labelling compactifications . One has
\begin{equation}\label{stoz}
    \frac{r^2}{1+r^2}(y^0)^2 - (y^1)^2- (y^2)^2 =0.
\end{equation}
Eq. (\ref{stoz}) presents two cones with a common vertex at
$\:(y^0,y^1,y^2)= (0,0,0)$. The induced metric on (\ref{stoz}) reads
\begin{equation}\label{line1}
    ds^2 =  - dt^2 +t^2 d\theta^2 .
\end{equation}
Generalization of the 2d CM space to the Nd spacetime  has the form
\begin{equation}\label{line2}
ds^2 = -dt^2 +dx^k dx_k +  t^2 d\theta^2,
\end{equation}
where $t,x^k \in \mathbb{R}^1,~\theta\in \mathbb{S}^1~(k= 1,\ldots,
N-2)$. One term in the metric (\ref{line2}) disappears/appears at
$t=0$, thus the CM space may be used to model the big-crunch/big-bang
type singularity \cite{Khoury:2001bz}. In what follows we restrict
our considerations to the 2d CM space. Later, we make comments
concerning generalizations.

It is clear that the CM space  is locally isometric to the Minkowski
space at each point except the vertex $t=0$. The CM space is not a
manifold, but an orbifold due to this vertex. The Riemann tensor
components vanish for $t\neq 0$ and cannot be defined at $t=0$, since
one dimension disappears/appears there. There is a space-like
singularity at $t=0$ of removable type because any time-like geodesic
with $t<0$ can be extended to some time-like geodesic with $t>0$.
However, such an extension cannot be unique due to the Cauchy problem
for the geodesic equation at the vertex (compact dimension shrinks
away and reappears at $t=0$).

\subsection{The RCM space}

Since trajectory of a \textit{test} particle coincides (by
definition) with time-like geodesic, there is no obstacle for the
test particle to reach and leave the CS. However, the Cauchy problem
for a geodesic equation at the CS is not well defined.  As the
result, a test particle `does not know where to go' at the
singularity. Thus, the singularity acts as `generator' of uncertainty
in the propagation of a test particle from the pre-singularity to
post-singularity era. In the present paper we propose to solve this
problem by replacement of a test particle by a \textit{physical} one.
The test and physical particles differ in a number of ways. For
instance, physical particle's own gravitational field effects its
motion \cite{Poisson:2004gg} and may modify the singularity of the CM
space. We assume that  these effects may be modelled by replacing the
CM space by a regularized compactified Milne, RCM, space in such a
way that the big-crunch/big-bang type singularity of the CM space is
replaced by the big-bounce type singularity.  In the RCM space the
Cauchy problem  does not occur because compact space dimension does
not contract to a point, but to some `small' value. As the result the
propagation of a particle is uniquely defined in the entire
spacetime. Particle's propagation in the RCM space is similar to the
corresponding one in the de Sitter space \cite{WP,Piechocki:2003hh}.

We define the RCM space by the following embedding into the 3d
Minkowski space
\begin{equation}\label{rem}
    y^0(t,\theta) = t\sqrt{1+r^2},~~~y^1(t,\theta) =
    r\sqrt{t^2+\epsilon^2}\sin(\theta/r),~~~y^2(t,\theta) = r\sqrt{t^2+
    \epsilon^2}\cos(\theta/r),
\end{equation}
and we have the relation
\begin{equation}\label{conr}
\frac{r^2}{1+r^2}(y^0)^2 - (y^1)^2- (y^2)^2 =-\epsilon^2r^2 .
\end{equation}

The induced metric on the RCM space reads
\begin{equation}\label{line3}
ds^2_\epsilon =  - (1+\frac{r^2\epsilon^2}{t^2+\epsilon^2})\;dt^2
+(t^2+\epsilon^2)\; d\theta^2 ,
\end{equation}
where $\epsilon\in \dR$ is a small number. It is clear that now the
space dimension $\theta$ does not shrink to zero at $t=0$. The scalar
curvature has the form
\begin{equation}\label{sca}
\mathcal{R}_\epsilon=
\frac{2\epsilon^2(1+r^2)}{(\epsilon^2(1+r^2)+t^2)^2}
\end{equation}
and the Einstein tensor corresponding to the metric (\ref{line3}) is
zero, thus (\ref{line3}) defines some vacuum solution to the 2d
Einstein equation.

It is evident that at $t\neq 0$ we have
\begin{equation}\label{comp}
\lim_{\epsilon\to 0} ds^2_\epsilon =  - dt^2
+t^2d\theta^2~~~~~\mathrm{and}~~~~~\lim_{\epsilon\to 0}
\mathcal{R}_\epsilon = 0.
\end{equation}
It is obvious that (\ref{conr}) turns into (\ref{stoz}) as
$\epsilon\rightarrow 0$.

\begin{figure}[h]
\vspace{-0.02\textwidth} \hspace{-0.02\textwidth}
\includegraphics[width=0.55\textwidth]{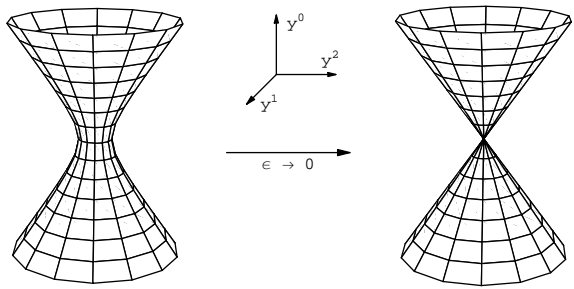}
\vspace{-0.05\textwidth} \caption{Embeddings of RCM and CM spaces.}
\end{figure}

Figure 1  presents  the RCM and CM spaces embedded  into the 3d
Minkowski space. We can see that the big-crunch/big-bang singularity
of the CM space is represented in the RCM by the big-bounce type
singularity.

\section{Classical dynamics}
An action integral, $\mathcal{A}$, describing a relativistic test
particle of mass $m$ in a gravitational field $g_{kl},~(k,l=0,1)$ may
be defined by
\begin{equation}\label{action}
\mathcal{A}=\int d\tau\:
L(\tau),~~~~ L(\tau):=\frac{m}{2}\:(\frac{\dot{x}^k \dot{x}^l}{e}
g_{kl}-e),
\end{equation}
where $\dot{x}^k :=dx^k/d\tau,~\tau$ is an evolution parameter,
$e(\tau)$ denotes the `einbein' on the world-line, $x^0$ and $x^1$
are time and space coordinates, respectively.

\noindent In case of the CM and RCM spaces the Lagrangian $L_\epsilon
$ reads
\begin{equation}\label{lag}
    L_\epsilon(\tau)=  \frac{m}{2e}\:\big((t^2+\epsilon^2) \dot{\theta}^2 -
    (1+\frac{r^2\epsilon^2}{t^2+\epsilon^2})\dot{t}^2 -e^2\big),
\end{equation}
where $\epsilon =0$ corresponds to the CM space. The action
(\ref{action}) is invariant under reparametrization with respect to
$\tau$. This gauge symmetry leads to the constraint
\begin{equation}\label{con}
    \Phi_\epsilon :=  p_k p_l \;g^{kl} + m^2 = \frac{p^2_\theta}{(t^2 + \epsilon^2)}-
    \frac{p^2_t}{1+\frac{r^2\epsilon^2}{t^2+\epsilon^2}} + m^2 =0,
\end{equation}
where  $p_t := \partial L_\epsilon/\partial\dot{t}\:$ and $p_\theta
:=\partial L_\epsilon/\partial\dot{\theta}\:$ are  canonical momenta,
and where $g^{kl}$ denotes an inverse of the metric $g_{kl}$ defined
by the line element (\ref{line3})  (case $\epsilon =0$ corresponds to
the CM space).

Variational principle applied to (\ref{action}) gives  equations of
motion of a particle
\begin{equation}\label{eq}
\frac{d}{d\tau} p_\theta = 0, ~~~\frac{d}{d\tau}p_t - \frac{\partial
L}{\partial t} = 0,~~~\frac{\partial L}{\partial e} = 0.
\end{equation}

Since during evolution of the system $p_\theta$ is conserved, due to
(\ref{eq}), we can analyze the behaviour of $p_t$ by making use of
the constraint (\ref{con}). In case of the CM space $(\epsilon =0)$,
for $p_\theta\neq 0$  there must be $p_t \rightarrow\infty$ as
$t\rightarrow 0$. This problem cannot be avoided by different choice
of coordinates\footnote{The system of coordinates we use, $
(t,\theta)\in \dR^1 \times \dS^1 $, is  natural for the spacetimes
with the topologies presented in Fig. 1.}.  It is connected with the
vanishing/appearance of the space dimension $\theta$ at $t=0$.
Another interpretation of this problem is that different geodesics
cross each other with the relative speed reaching the speed of light
as they approach the singularity at $t=0$.

The dynamics of a physical particle in the RCM space $(\epsilon \neq
0)$ does not suffer from such a problem, since for $p_\theta\neq 0$
the momentum component $p_t$ does not need to `blow up' to satisfy
(\ref{con}).

\subsection{Geodesics in CM and RCM spaces}

It was found in \cite{Malkiewicz:2005ii} an analytic general solution
to (\ref{eq}), for $\epsilon =0$, in the form
\begin{equation}\label{sol1}
    \theta(t)= \theta_0  - \sinh^{-1} \Big(\frac{p_\theta}{mt}\Big),
\end{equation}
where $(p_\theta, \theta_0) \in \dR^1\times\dS^1$. It is clear
that geodesics (\ref{sol1}) `blow up' at $t=0$, which is
visualized in Fig. 2.

For $\epsilon\neq 0$, Eqs. (\ref{eq})  read
\begin{equation}\label{eq2}
\frac{m(t^2+\epsilon^2)\dot{\theta}}{e}=p_\theta=const,~~~~
e^2=(1+\frac{r^2\epsilon^2}{t^2+\epsilon^2})\dot{t}^2-(t^2+
\epsilon^2)\dot{\theta}^2
\end{equation}
and
\begin{equation}\label{eq3}
\big(1+\frac{r^2\epsilon^2}{t^2+\epsilon^2}\big)\ddot{t}-
\big(1+\frac{r^2\epsilon^2}{t^2+\epsilon^2}\big)
\big(\frac{\dot{e}}{e}\big)\dot{t}-  \frac{r^2\epsilon^2t}
{(t^2+\epsilon^2)^2}\dot{t}+\dot{\theta}^2t=0.
\end{equation}
From  (\ref{eq2}) and (\ref{eq3}) we get
\begin{equation}\label{sol2}
\Big(\frac{d\theta}{dt}\Big)^2=\frac{p_\theta^2(1+\frac{r^2\epsilon^2}{{t}^2+\epsilon^2})}
{m^2(t^2+\epsilon^2)^2+p_\theta^2(t^2+\epsilon^2)},
\end{equation}
where $p_\theta \in \dR^1$. General solution to (\ref{sol2}) reads
\begin{equation}\label{sol3}
\theta (t) = \theta_0 ~+ ~p_\theta \int_{-\infty}^{t}
\,d{\tau}\sqrt{\frac{1+\frac{r^2\epsilon^2}{{\tau}^2+\epsilon^2}}{m^2({\tau}^2+\epsilon^2)^2+
p_\theta^2({\tau}^2+\epsilon^2)}}
\end{equation}
where $\theta_0\in \dS^1$. The integral in (\ref{sol3}) cannot be
calculated analytically. Numerical solution of  (\ref{sol2}) is
presented in Fig. 2.

\begin{figure}[h]
\hspace{-0.18\textwidth}
\begin{minipage}[b]{0.45\textwidth}
\flushleft
\includegraphics[width=0.56\textwidth,height=0.42\textwidth]{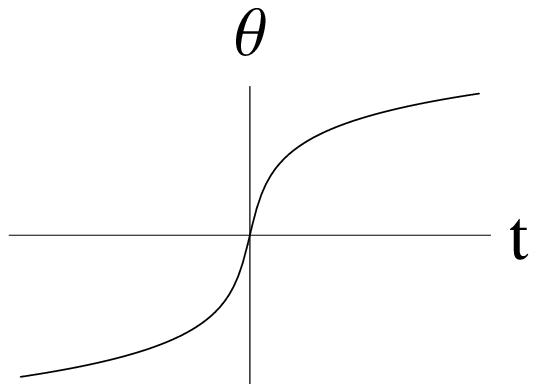}
\nolinebreak
\begin{minipage}[b]{0.22\textwidth}
\includegraphics[width=\textwidth]{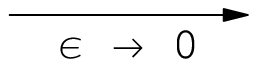}
\vspace{0.11\textwidth}
\end{minipage}
\nolinebreak
\includegraphics[width=0.56\textwidth,height=0.42\textwidth]{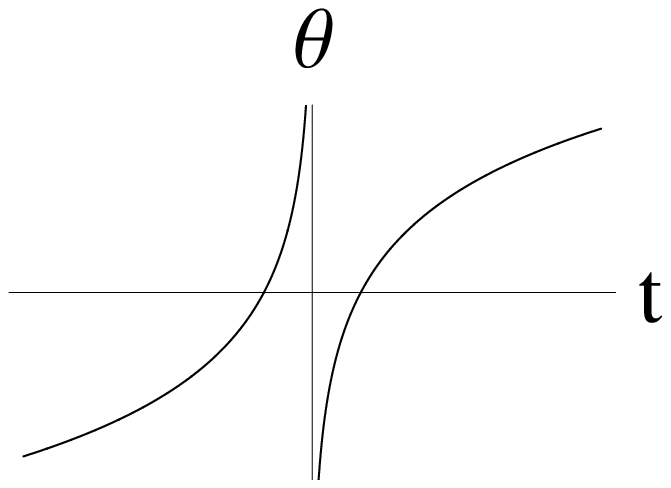}
\end{minipage}
\caption{Geodesics in RCM space (the left graph) and CM space (the
right graph).}
\end{figure}

Figure 2 shows  that a geodesic in the RCM space  is bounded and
continuous in the neighborhood of the singularity. In contrary, a
geodesic in the CM space (drawing by making use of (\ref{sol1}))
blows up as $t\rightarrow\pm 0$.

\subsection{Phase space and basic observables}

We define a phase space  to be the set of independent parameters
(variables) defining all particle geodesics. Thus the pase space,
$\Gamma$, for (\ref{sol3}) reads
\begin{equation}\label{phase1}
\Gamma := \{(\sigma,p_\sigma)\;|\; \sigma\in \dR^1\; mod\;2\pi r,\;
p_\sigma \in \dR^1\}= \dS^1 \times \dR^1.
\end{equation}
The Cauchy problem at the singularity results from the
vanishing/appearance of the space dimension $\theta$ at $t=0$. It is
fairly probable  that {\it any}  simple regularization of the
singularity of the CM space that prevents such collapse will lead to
the cylindrical phase space (\ref{phase1}).

In   \cite{Malkiewicz:2005ii}  we have analyzed four types of
propagations of a particle in the CM space. Now we can see that the
regularization prefers the propagation in the CM space  of the de
Sitter type (see, Sec. III D of \cite{Malkiewicz:2005ii}), because
only in this case the phase space topology has the form
(\ref{phase1}).

Now, let us identify the \textit{basic} canonical functions on the
phase space, i.e. observables that can be used to define any
\textit{composite} observable of the underlying classical system. In
case a phase space includes a variable with non-trivial topology,
i.e. different from $\dR^1$, it is a serious problem. However, it has
been solved in two (equivalent) ways not long ago. In what follows we
use the method used in the group theoretical quantization (see,
\cite{Kastrup:2005xb} and references therein). In the next section we
explain relation with another method.

A natural choice  \cite{Kastrup:2005xb} of the basic functions on
(\ref{phase1})  is

\begin{equation}\label{bf}
S:=\sin(\sigma/r),~~C:=\cos(\sigma/r),~~P:=r p_\sigma .
\end{equation}
The basic observables  $S$ and $C$ are smooth single-valued functions
on $\dS^1$ (contrary to $\sigma$). The observables (\ref{bf}) satisfy
the Euclidean algebra $e(2)$ on $\Gamma$
\begin{equation}\label{n1}
    \{S,C\}= 0,~~\{P,S\}=C,~~\{P,C\}=-S,
\end{equation}
where
\begin{equation}\label{pb}
\{\cdot,\cdot\} := \frac{\partial~\cdot}{\partial
p_\sigma}\frac{\partial~\cdot}{\partial\sigma} -
\frac{\partial~\cdot}{\partial \sigma}\frac{\partial~\cdot}{\partial
p_\sigma}.
\end{equation}
It is shown in \cite{Kastrup:2005xb} that the Euclidean group $E(2)$
can be used as the canonical group \cite{CJ} of  the phase space
$\Gamma$.

\section{Quantization of  phase space}

By quantization we mean  finding  an irreducible unitary
representation of  the symmetry group  of the phase space of the
underlying classical system.

The group $E(2)$ has the following irreducible unitary representation
\cite{Kastrup:2005xb}
\begin{equation}\label{ua}
    [U(\alpha)\psi](\beta):= \psi[(\beta -
    \alpha)\;mod\;2\pi],~~~~\mathrm{for\;
    rotations}~~~~z\rightarrow e^{i\alpha}z,
\end{equation}
\begin{equation}\label{ut}
    [U(t)\psi](\beta):= [\exp{-i(a\cos\beta + b\sin\beta)}]\psi(\beta),~~~~
    \mathrm{for\;translations}~~~~ z\rightarrow z+t,
\end{equation}
where $z=|z|\;e^{i\beta},~~t=a+bi$, and where $\psi\in L^2(\dS^1)$.

Making use of the Stone theorem, we can find an (essentially)
self-adjoint representation of the algebra (\ref{n1}). One has
\begin{equation}\label{algebra}
    [\hat{C},\hat{S}]=0,~~[\hat{P},\hat{S}]=-i\hat{C},
       ~~[\hat{P},\hat{C}]=i\hat{S},
\end{equation}
where
\begin{equation}\label{mom}
    \hat{P}\varphi(\beta):=
    -i\frac{\partial}{\partial\beta}\varphi(\beta),~~~~
    \hat{S}\varphi(\beta):=\sin\beta\;\varphi(\beta),~~~~
    \hat{C}\varphi(\beta):=\cos\beta\;\varphi(\beta).
\end{equation}
The domain, $\Omega_\lambda$, of operators $\hat{P}, \hat{S}, \hat
{C}$ reads
\begin{equation}\label{domain}
\Omega_\lambda := \{\varphi\in L^2(\dS^1)~|~\varphi\in
C^{\infty}[0,2\pi],\;\varphi^{(n)} (2\pi) = e^{i\lambda}\varphi
^{(n)}(0),~~n=0,1,2,\dots\},
\end{equation}
where $ 0\leq\lambda < 2\pi $ labels various representations of
$e(2)$ algebra. The space $\Omega_\lambda$ is dense in $L^2(\dS^1)$
so the unbounded operator $\hat{P}$ is well defined. As the operators
$\hat{S}$ and $\hat{C}$ are bounded on the entire $L^2(\dS^1)$, the
space $\Omega_\lambda$ is a common invariant domain for all operators
and their products.

In \cite{Malkiewicz:2005ii} we have found that the representation
of the algebra specific to the case considered there in Sec. III
D, has the form
\begin{equation}\label{quant3}
[\hat{\alpha},\hat{U}]= \hat{U},
\end{equation}
with
\begin{equation}\label{quant}
\hat{\alpha}\varphi(\beta):= -i\frac{d}{d\beta}\varphi(\beta),~~~~~~
\hat{U}\varphi(\beta) :=e^{i\beta}\varphi(\beta),
\end{equation}
where $0\leq\beta<2\pi$ and $\varphi\in \Omega_\lambda$. However,
both representations, (\ref{mom}) and (\ref{quant}), are in fact the
same owing to
\begin{equation}\label{same}
 e^{i\beta} = \cos\beta + i \sin\beta,~~~~~[\cos\beta, \sin\beta]=0 .
\end{equation}

The space $\Omega_\lambda$, where $0\leq\lambda < 2\pi$, may be
spanned by the set of orthonormal eigenfunctions of the operator
$\hat{\alpha}$
\begin{equation}\label{quant4}
f_{k,\lambda}(\beta):= (2\pi)^{-1/2}\exp{i\beta
(k+\lambda/2\pi}),~~~~~k=0,\pm 1,\pm 2,\ldots
\end{equation}
However, the functions (\ref{quant4}) are {\it continuous} on $\dS^1$
only in the case when  $\lambda =0$. Thus, the requirement of the
continuity removes the ambiguity of quantization.

\section{Stability of System}

To examine the stability problem of our system, we use the
Hamiltonian formulation of the dynamics of a particle.

\subsection{Classical level}

By stability of the dynamics of a \textit{classical} particle we mean
such an evolution of a particle that can be described by the
canonical variables which are bounded and continuous functions.

Direct application of the results of \cite{Turok:2004gb} gives the
following expression for the extended Hamiltonian \cite{PAM, MHT},
$H_\epsilon$, of a particle
\begin{equation}\label{ham1}
    H_\epsilon = \frac{1}{2}\; C_\epsilon \;\Phi_\epsilon ,
\end{equation}
where $C_\epsilon$ is an arbitrary function of an evolution parameter
$\tau$, and where $\Phi_\epsilon$ is the first-class constraint
defined by (\ref{con}). The equations of motion for canonical
variables $(t,\theta;p_t,p_\theta)$ read
\begin{equation}\label{e1}
    \frac{dt}{d\tau}=\{t,H_\epsilon\}=  -C_\epsilon(\tau)\frac{p_t (t^2 +\epsilon^2)}
    {t^2 +\epsilon^2+r^2 \epsilon^2},
\end{equation}
\begin{equation}\label{e2}
\frac{d\theta}{d\tau}=\{\theta,H_\epsilon\}=C_\epsilon
(\tau)\frac{p_\theta}{t^2 +\epsilon^2},
\end{equation}
\begin{equation}\label{e3}
\frac{dp_t}{d\tau}=\{p_t,H_\epsilon\}=
C_\epsilon(\tau)\Big(\frac{t p_\theta^2}{(t^2 +\epsilon^2)^2}
+\frac{t p^2_t\epsilon^2 r^2}{(t^2+\epsilon^2+r^2\epsilon^2)^2}
\Big),
\end{equation}
\begin{equation}\label{e4}
\frac{dp_\theta}{d\tau}=\{p_\theta,H_\epsilon\}= 0,
\end{equation}
where
\begin{equation}
    \{\cdot,\cdot\}=\frac{\partial\cdot}{\partial t}\frac{\partial\cdot}
    {\partial p_t} - \frac{\partial\cdot}{\partial p_t}\frac{\partial\cdot}
    {\partial t} +   \frac{\partial\cdot}{\partial \theta}\frac{\partial\cdot}
    {\partial p_\theta} - \frac{\partial\cdot}{\partial p_\theta}\frac{\partial\cdot}
    {\partial \theta}  \nonumber
\end{equation}
To solve (\ref{e1})-(\ref{e4}), we use the gauge  $\tau = t$. In this
gauge (\ref{e1}) leads to
\begin{equation}\label{e5}
    C_\epsilon(t)= -\frac{t^2+\epsilon^2+r^2\epsilon^2}{p_t(t^2+\epsilon^2)}.
\end{equation}
Insertion of (\ref{e5}) into (\ref{e3}) and taking into account
(\ref{con}) gives
\begin{equation}\label{e6}
   \frac{d}{dt}p_t^2 = -
    \frac{2tp_t^2r^2\epsilon^2}{(t^2+\epsilon^2)(t^2+\epsilon^2+r^2\epsilon^2)} -
    \frac{2tp_{\theta}^2(t^2+\epsilon^2+r^2\epsilon^2)}
    {(t^2+\epsilon^2)^3}.
\end{equation}
Solution to (\ref{e6}) reads
\begin{equation}\label{e7}
    p_t^2 =c_1\Big( \frac{c_1p_{\theta}^2}{t^2+\epsilon^2}+c_2\Big)\;
    \frac{t^2+\epsilon^2+r^2\epsilon^2}{t^2+\epsilon^2},
\end{equation}
where $p_\theta$ does not depend on time due to (\ref{e4}). This
result is consistent with the constraint (\ref{con}) if we put
$c_1=1$ and $c_2=m^2$. Insertion of (\ref{e7}) into (\ref{e5}) yields
an explicit expression for $C_\epsilon(t)$. Next, insertion of so
obtained $C_\epsilon(t)$ into (\ref{e2}) gives (\ref{sol2}) with the
solution (\ref{sol3}). Thus, we have found complete
solutions\footnote{This way we have also verified, in the gauge $\tau
=t$, the equivalence between our Hamiltonian and the Lagrangian
formulations of the dynamics of a particle.} to the equations
(\ref{e1})-(\ref{e4}).

It results from the functional form of solutions that for
$\epsilon\neq 0$ the propagation of canonical variables is regular,
i.e. has no singularities for any value of time. One may also verify
that the constraint equation (\ref{con}), with $p_t$ determined by
(\ref{e7}), is satisfied for each value of time either. Since
$C_\epsilon(t)$, determined by (\ref{e5}) is bounded, the Hamiltonian
$H_\epsilon$ defined by (\ref{ham1}) is weakly zero  due to the
constraint (\ref{con}).

Now, let us analyze the case $\epsilon =0$, which corresponds to the
evolution of a particle in the CM space. The solutions of
(\ref{e1})-(\ref{e4}), in the gauge $\tau =t$, are the following
\begin{equation}\label{ee1}
    C_0(t) = -1/p_t ,
\end{equation}
\begin{equation}\label{ee2}
    p_t^2 = p_\theta^2 /t^2 +m^2
\end{equation}
\begin{equation}\label{ee3}
    \theta(t) = - sinh^{-1} \Big(p_\theta /mt\Big)+
    const ,
\end{equation}
(where $p_\theta = const$), in agreement with (\ref{sol1}). The
constraint equation (\ref{con}) reads
\begin{equation}\label{ee4}
    \Phi_0 = p_\theta^2 /t^2 - p_t^2 + m^2 = 0.
\end{equation}

It results from (\ref{ee2}) and (\ref{ee3}) that only for $p_\theta
=0$ the propagation of a particle in the CM space is regular for any
value of time. The Hamiltonian (\ref{ham1}) is also regular and  is
weakly equal to zero. Quite different situation occurs in the case
$p_\theta\neq 0$. The equations (\ref{ee2})-(\ref{ee4}) and the
Hamiltonian are singular at $t=0$. Thus the dynamics of a particle is
unstable\footnote{The division of the set of geodesics into regular
and unstable depends on the choice of coordinates, but it always
includes  the unstable ones.}.

\subsection{Quantum level}

By stability of dynamics of a \textit{quantum} particle we mean the
boundedness from below of its quantum Hamiltonian.

To construct the quantum Hamiltonian of a particle we use  the
following mapping (see, e.g. \cite{Ryan:2004vv})
\begin{equation}\label{mapp}
    p_k p_l g^{kl} \longrightarrow \Box  :=
    (-g)^{-1/2}\partial_k [(-g)^{1/2} g^{kl} \partial_l ],
\end{equation}
where $g:=det [g_{kl}]$ and $\partial_k := \partial/\partial x^k$.
The Laplace-Beltrami operator, $\Box$,  is invariant under the change
of spacetime coordinates and it leads to  Hamiltonians that give
results consistent with  experiments  \cite{Ryan:2004vv}, and which
has been used in theoretical cosmology (see, \cite{Turok:2004gb} and
references therein).

In the case of the CM space   the Hamiltonian,  for $t<0$ or $t>0$,
reads
\begin{equation}\label{nh}
\hat{H} = \Box + m^2 =  \frac{\partial}{\partial t^2} +
      \frac{1}{t}\frac{\partial}{\partial t} - \frac{1}{t^2}\frac{\partial^2}
      {\partial \theta^2} + m^2 .
\end{equation}
The operator $\hat{H}$ was obtained by making use of (\ref{mapp}),
(\ref{line1}) and (\ref{ham1}) in the gauge $C_\epsilon =2$ (for
$\epsilon =0$). In this gauge\footnote{In the preceding subsection,
concerning the classical dynamics, the choice of gauge was different.
But since the theory we use is gauge invariant, the different choice
of the gauge does not effect physical results.} the Hamiltonian
equals the first class constraint (\ref{ee4}). Thus the Dirac
quantization scheme \cite{PAM,MHT} leads to the equation
\begin{equation}\label{hcn}
\hat{H} \psi(\theta,t)= 0.
\end{equation}
The space of solutions to (\ref{hcn}) defines the domain of
boundedness of $\hat{H}$ from below (and from above).

Let us find the non-zero solutions of (\ref{hcn}). Separating the
variables
\begin{equation}\label{sep}
  \psi(\theta,t):= A(\theta)\;B(t)
\end{equation}
leads to the equations
\begin{equation}\label{eqth}
 d^2 A/d\theta^2 + \rho^2 A = 0,~~~~\rho\in
\dR
\end{equation}
and
\begin{equation}\label{eqt}
    \frac{d^2 B}{d t^2}+ \frac{1}{t} \:\frac{dB}{dt} + \frac{m^2 t^2 +\rho ^2}{t^2}\;B =
    0,~~~~t\neq 0,
\end{equation}
where $\rho$ is a constant of separation. Two independent continuous
solutions  on  $\dS^1$ read
\begin{equation}\label{solth}
     A_1(\rho,\theta)= a_1 \cos(\rho\theta),~~~~A_2(\rho,\theta)= a_2
     \sin(\rho\theta),~~~~~~~a_1, a_2 \in \dR .
\end{equation}
Two independent solutions  on $ \dR $ (for $t<0$ or $t>0$) have the
form  \cite{Arfken:2005,SWM}
\begin{equation}\label{solt}
  B_1(\rho,t)= b_1  \Re J(i\rho,mt),~~~~B_2(\rho,t)= b_2 \Re Y(i\rho,mt),~~~~~~~b_1,
  b_2 \in \dC ,
\end{equation}
where $\Re J$ and $\Re Y$ are  the real parts of Bessel's and
Neumann's functions, respectively. Since $\rho\in\dR$, the number of
independent solutions is: $2 \times 2 \times \infty$ ( for $t<0$ and
$t>0$).

At this stage we define the scalar product on the space of solutions
(\ref{solth}) and (\ref{solt}) as follows
\begin{equation}\label{scalar}
    <\psi_1|\psi_2> := \int_{\widetilde{\Gamma}} d \mu \;\overline{\psi}_1 \;\psi_2,~~~~~~d\mu
    :=\sqrt{-g}\; d\theta \;dt = |t|\; d\theta \;dt,
\end{equation}
where $\widetilde{\Gamma}:= [-T,0[ \times \dS^1$ (with $T
>0$) in the pre-singulaity epoch, and $\widetilde{\Gamma}:= ]0,T]
\times \dS^1$  in the post-singularity epoch. We assume that the CM
space can be used to model the universe only  during its quantum
phase, which lasts the period $[-T, T$].

Now we construct an orthonormal basis, in the left neighborhood of
the cosmological singularity,  out of the solutions (\ref{solth})
and (\ref{solt}). One can  verify that the solutions (\ref{solth})
are orthonormal and continuous on $\dS^1$ if $\;a_1  = \pi^{-1/2}=
a_2\;$ and $\;r\rho = 0,\pm 1,\pm 2,\ldots$. (This set of
functions coincides with the basis (\ref{quant4}) that spans the
subspace $\Omega_\lambda$ if we replace  $k$ by $r\rho$.) Some
effort is needed to construct the set of orthonormal functions out
of $\Re J(i\rho,mt)$ and $\Re Y(i\rho,mt)$. First, one may verify
that these functions are square-integrable on the interval
$[-T,T]$. This is due to the choice of the measure in the scalar
product (\ref{scalar}), which leads to the boundedness of the
corresponding integrants. Second, having normalizable set of four
independent functions, for each $\rho$, we can turn it into an
orthonormal set by making use of the Gram-Schmidt procedure (see,
e.g. \cite{Arfken:2005}). Our orthonormal and countable set of
functions may be used to define the span $\mathcal{F}$. The
completion of $\mathcal{F}$ in the norm induced by the scalar
product (\ref{scalar}) defines the Hilbert spaces
$L^2(\widetilde{\Gamma} \times \dS^1,d\mu)$. It is clear that the
same procedure applies to the right neighborhood of the
singularity.

\begin{figure}[h]
\centering \subfigure[\ near the singularity]{

\includegraphics[width=2.3in]{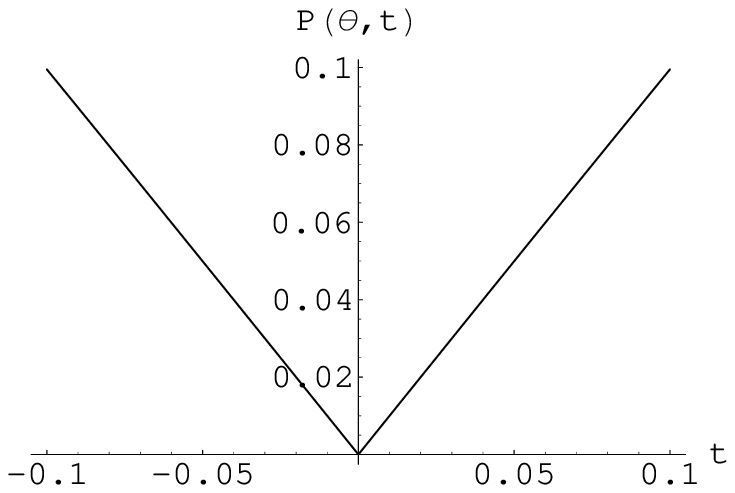}}
\hspace{0.4in} \subfigure[\ on large scale]{

\includegraphics[width=2.3in]{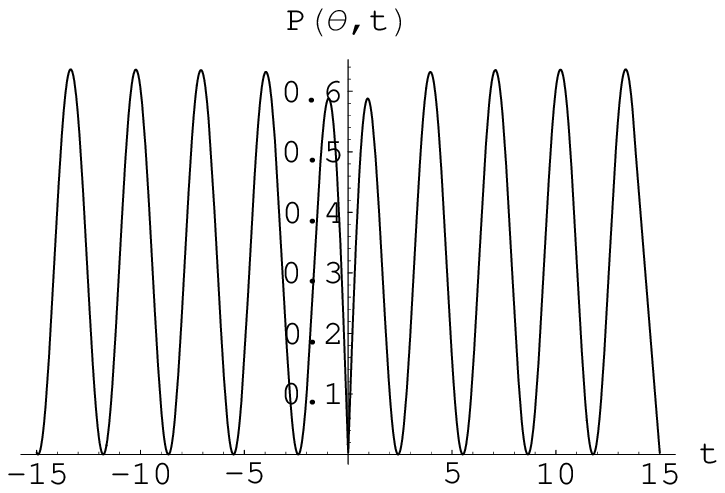}}
\caption{Probability density corresponding to $\psi(\theta,t)=
A_1(0,\theta)\;\Re J(0,t)$}

\end{figure}

\begin{figure}[h]
\centering \subfigure[\ near the singularity]{

\includegraphics[width=2.3in]{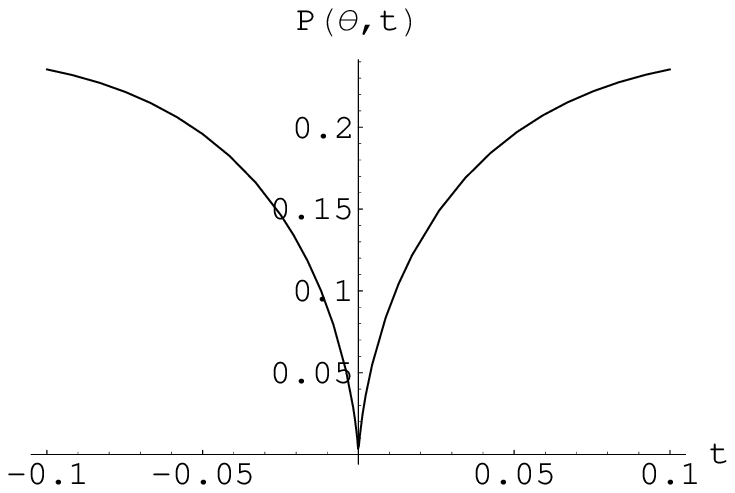}}
\hspace{0.4in} \subfigure[\ on large scale]{

\includegraphics[width=2.3in]{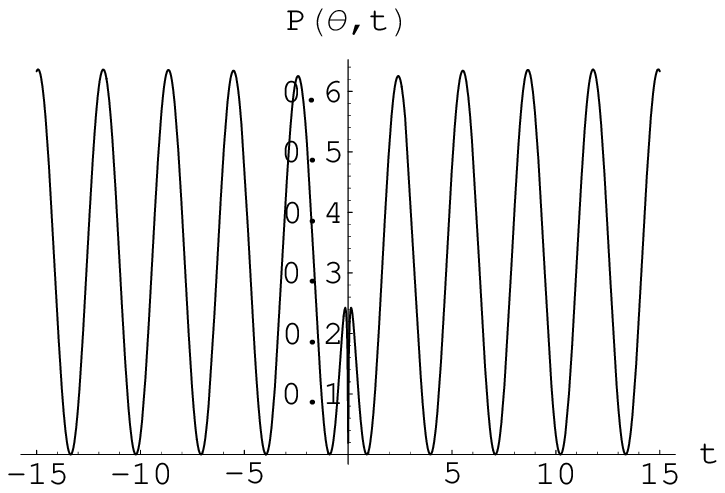}}
\caption{Probability density corresponding to $\psi(\theta,t)=
A_1(0,\theta)\;\Re Y(0,t)$}
\end{figure}

Finally, we can prove that the Hamiltonian (\ref{nh}) is self-adjoint
on $L^2(\widetilde{\Gamma} \times \dS^1,d\mu)$. The proof is
immediate if we rewrite (\ref{hcn}) in the form
\begin{equation}\label{newf}
    \Box \;\psi = - m^2\;\psi .
\end{equation}
It is evident that on the orthonormal basis that we have constructed
above the operator $\Box$ is an identity operator multiplied by a
real constant $-m^2$.  The operator $\Box$ is bounded since
\begin{equation}\label{bound}
    \|\Box \| := \sup_{\|\psi\| =1}  \|\Box\;\psi\| = \sup_{\|\psi\| =1} \|
    -m^2 \; \psi\| =  m^2 < \infty ,
\end{equation}
where $\|\psi\|:=\sqrt{<\psi|\psi>}$. The operator $\Box$ is also
symmetric, because $m$ is a {\it real} constant. Since $\Box$ is
bounded and symmetric, it is  a self-adjoint operator (see, e.g.
\cite{MRS}).  Clearly, the self-adjointness of the Hamiltonian
(\ref{nh}) results from the self-adjointness of $\Box$.

We have constructed the two  Hilbert spaces: one for the
pre-singularity epoch, $\mathcal{H}^{(-)}$, and another one to
describe the post-singularity epoch, $\mathcal{H}^{(+)}$. Next
problem is to `glue' them into a single Hilbert space,
$\mathcal{H}=L^2([-T,T] \times \dS^1,d\mu)$, that is needed to
describe the entire quantum phase. From the mathematical point of
view the gluing seems to be problematic because the Cauchy problem
for the equation (\ref{hcn}) is not well defined\footnote{Except one
case discussed later.} at $t=0$, and because we have assumed that
$t\neq 0$ in the process of separation of variables to get Eqs.
(\ref{eqth}) and (\ref{eqt}). However, arguing based on the physics
of the problem enables the gluing. First of all we have already
agreed that a \textit{classical} test particle is able to go across
the singularity (see, subsection II B). One can also verify that the
probability density
\begin{equation}\label{amp}
    P(\theta,t):= \sqrt{-g}\;|\psi(\theta,t)|^2 = |t|\;|\psi(\theta,t)|^2
\end{equation}
is bounded  and continuous in the domain $\;[-T,T] \times \dS^1$.
Figures 3 and 4 illustrate the behavior of $P(\theta,t)$ for two
examples of gluing the  solutions having $\rho =0$. The cases with
$\rho \neq 0$ have similar properties. Thus, the assumption that the
gluing is possible is justified. However one can glue the two Hilbert
spaces in more than one way,  as it was done in the quantization of
the phase space in our previous paper \cite{Malkiewicz:2005ii}. In
what follows we present  two cases, which are radically different.

\subsubsection{Deterministic propagation}

Among all solutions (\ref{solt}) there is one, corresponding to $\rho
=0$, that attracts an attention \cite{SWM}. It reads
\begin{equation}\label{n1B}
    B_1(0,mt)= b_1\;\Re J(0,mt),~~~~~~b_1\in \dR ,
\end{equation}
and has the following power series expansion close to $t=0$
\begin{equation}\label{psn}
   B_1(0,x)/b_1 = 1- \frac{x^2}{4}+\frac{x^4}{64} - \frac{x^6}{2304}
   + \mathcal{O}[x^8] .
\end{equation}
It is visualized in Fig. 5a. The solution (\ref{n1B}) is smooth at
the singularity, in spite of the fact that (\ref{eqt}) is singular at
$t=0$.

\begin{figure}[h]
\centering \subfigure[]{

\includegraphics[width=2.3in]{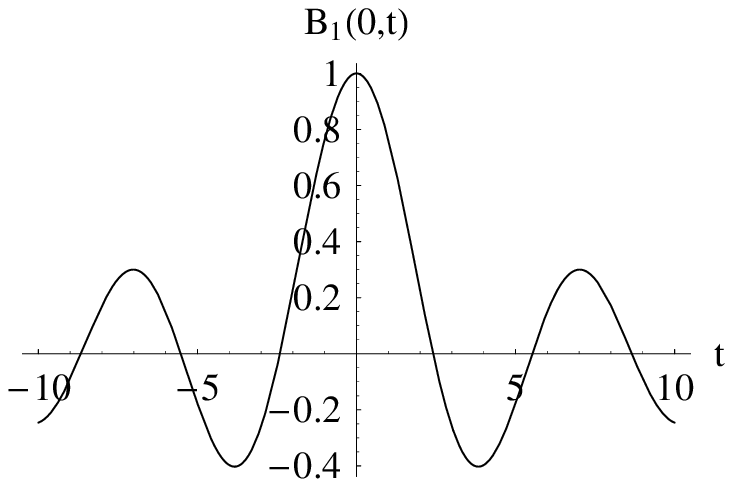}}
\hspace{0.4in} \subfigure[]{

\includegraphics[width=2.3in]{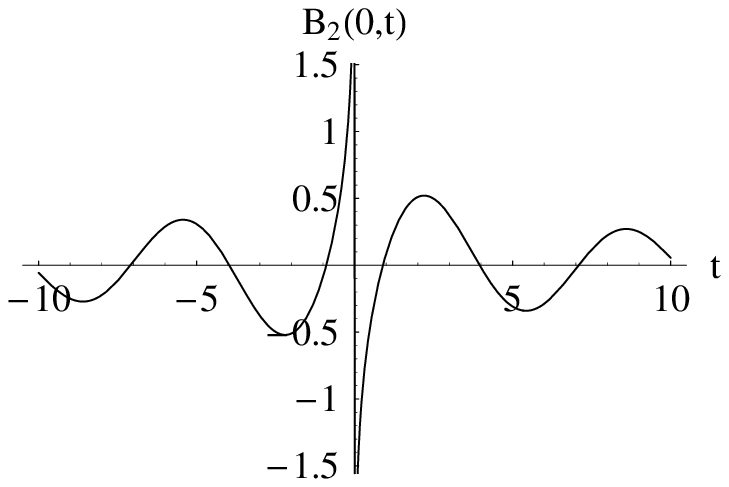}}
\caption{Continuous (a) and singular (b) propagations of a particle
with $\rho =0$. }
\end{figure}

It defines a solution to (\ref{hcn}) that does not depend on
$\theta$, since the non-zero solution (\ref{solth}) with $\rho=0$ is
just a constant. Thus, it is unsensitive to the problem that one
cannot choose a common coordinate system for both $t<0$ and $t>0$.

The solution $B_1$ (and the trivial solution $B_0 := 0$) can be used
to construct a one-dimensional Hilbert space $\mathcal{H}=L^2([-T,T]
\times \dS^1,d\mu)$. The scalar product is defined by (\ref{scalar})
with $\widetilde{\Gamma}$ replaced by $\Gamma := [-T,T] \times
\dS^1$. It is obvious that the Hamiltonian is self-adjoint on
$\mathcal{H}$.

The solution (\ref{n1B}) is {\it continuous} at the singularity. It
describes an unambiguous propagation of a quantum particle. Thus, we
call it a {\it deterministic} propagation. It is similar to the
propagation of a particle in the RCM space considered in the next
subsection.

Since (\ref{eqt}) is a second order differential equation, it should
have two independent solutions. However, the second solution cannot
be continuous at $t=0$. One may argue as follows: The solution
(\ref{n1B}) may be obtained by ignoring the restriction $t \neq 0$
and solving (\ref{eqt}) with the following initial conditions
\begin{equation}\label{incon}
    B(0,0)=1,~~~~~~dB(0,0)/dt = 0.
\end{equation}
Equations (\ref{eqt}) and (\ref{incon}) are consistent, because the
middle term of the r.h.s. of (\ref{eqt}) may be equal to zero due to
(\ref{incon}) so the resulting equation would be non-singular at
$t=0$. Another initial condition of the form $\;B(0,0)= const\;$ and
$\;dB(0,0)/dt = 0\;$ would be linearly dependent on (\ref{incon}).
Thus, it could not lead to the solution which would be continuous at
$t=0$ and linearly independent on (\ref{n1B}).

This qualitative reasoning can be replaced by a rigorous derivation
using the power series expansion method \cite{Arfken:2005}. Applying
this method one obtains that near the singularity $t=0$ the solution
to (\ref{eqt}) behaves like $\;t^\omega\;$ and that the corresponding
indicial equation reads
\begin{equation}\label{ind}
    \omega^2 = -\rho^2 .
\end{equation}
Thus, for $\rho\neq 0$  the two solutions behave like $\;t^{\pm i
\rho}\;$, i.e.  are bounded but not continuous (see, Eq.
(\ref{solt})). For $\rho =0$ the  indicial equation has only one
solution $\;\omega =0\;$ which leads to an analytic  solution to
(\ref{eqt}) defined by (\ref{n1B}). In such a case, it results from
the method of solving the singular linear second order equations
\cite{Arfken:2005}, the second solution to (\ref{eqt}) may behave
like $\;\ln |t|\;$. In fact it reads \cite{SWM}
\begin{equation}\label{n2B}
    B_2(0,mt)= b_2\;\Re Y(0,mt),~~~~~~b_2\in \dR ,
\end{equation}
and  is visualized in Fig. 5b. It cannot be called a deterministic
propagation due to the discontinuity  at the singularity $t=0$.

\subsubsection{Indeterministic propagation}

All solutions   (\ref{solt}), except (\ref{n1B}), are discontinuous
at $t=0$. This property is connected with the singularity of
(\ref{eqt}) at $t=0$. It is clear that due to such an obstacle the
identification of corresponding solutions on both sides of the
singularity is impossible. However there  are two natural
constructions of a Hilbert space out of $\mathcal{H}^{(-)}$ and
$\mathcal{H}^{(+)}$ which one can apply:\\
{\it (a) Tensor product of Hilbert spaces} \\
The Hilbert space is
defined in a standard way \cite{EP} as $\mathcal{H}:=
\mathcal{H}^{(-)}\otimes\mathcal{H}^{(+)}$ and it consists of
functions of the form
\begin{equation}\label{tp}
    f(t_1,\theta_1;t_2,\theta_2) \equiv (f^{(-)}\otimes f^{(+)})(t_1,\theta_1;t_2,\theta_2)
    := f^{(-)}(t_1,\theta_1)\;f^{(+)}(t_2,\theta_2) ,
\end{equation}
where $f^{(-)}\in \mathcal{H}^{(-)}$ and $f^{(+)}\in
\mathcal{H}^{(+)}$. The scalar product reads
\begin{equation}\label{sten}
    <f\;|\;g>:= <f^{(-)}|\;g^{(-)}>\;<f^{(+)}|\;g^{(+)}> ,
\end{equation}
where
\begin{equation}\label{stm}
  <f^{(-)}|\;g^{(-)}>:=\int_{-T}^0 dt_1 \int_0^{2\pi}d \theta_1
  \;|t_1|\; f^{(-)}(t_1,\theta_1)\; g^{(-)}(t_1,\theta_1)
\end{equation}
and
\begin{equation}\label{stp}
  <f^{(+)}|\;g^{(+)}>:=\int_0^{T} dt_2 \int_0^{2\pi}d \theta_2
  \;|t_2|\; f^{(+)}(t_2,\theta_2)\; g^{(+)}(t_2,\theta_2) .
\end{equation}
The action of the Hamiltonian is defined by
\begin{equation}\label{mamt}
  \hat{H} \big(f^{(-)}\otimes f^{(+)}\big):= \big(\hat{H} f^{(-)}\big)
  \otimes f^{(+)} +  f^{(-)}\otimes \big(\hat{H} f^{(+)}\big).
\end{equation}
The Hamiltonian is clearly self-adjoint on $\mathcal{H}$.

The quantum system described in this way appears to consist of two
independent parts. In fact it describes the same  quantum particle
but in two subsequent time intervals separated by the
singularity at $t=0$.\\
{\it (b) Direct sum of Hilbert spaces}\\
Another standard way \cite{EP} of defining the Hilbert space is
$\mathcal{H}:= \mathcal{H}^{(-)}\bigoplus\mathcal{H}^{(+)}$. The
scalar product reads
\begin{equation}\label{dssc}
    <f_1|f_2>:= <f_1^{(-)}|f_2^{(-)}> + <f_1^{(+)}|f_2^{(+)}> ,
\end{equation}
where
\begin{equation}\label{dsf}
    f_k := (f_k^{(-)},f_k^{(+)}) \in
    \mathcal{H}^{(-)}\times\mathcal{H}^{(+)},~~~~~~k=1,2,
\end{equation}
and where $f_k^{(-)}$ and $f_k^{(+)}$ are two completely independent
solutions in the pre-singularity and post-singularity epochs,
respectively. (The r.h.s of (\ref{dssc}) is defined by (\ref{stm})
and (\ref{stp}).)

The Hamiltonian action on $\mathcal{H}$ reads
\begin{equation}\label{hamds}
\mathcal{H}\ni (f^{(-)},f^{(+)})\longrightarrow \hat{H}
(f^{(-)},f^{(+)}):= (\hat{H} f^{(-)},\hat{H} f^{(+)}) \in
\mathcal{H}.
\end{equation}
It is obvious that $\hat{H}$ is self adjoint on $\mathcal{H}$.

By the construction, the space
$\mathcal{H}^{(-)}\bigoplus\mathcal{H}^{(+)}$ includes vectors like
$(f^{(-)},0)$ and $(0,f^{(+)})$, which give non-vanishing
contribution to (\ref{dssc}) (but yield zero in case (\ref{sten})).
The former state describes the annihilation of a particle at $t=0$.
The latter corresponds to the creation of a particle at the
singularity. These type of states do not describe the propagation of
a particle {\it across} the singularity. The annihilation/creation of
a massive particle would change the background. Such events should be
eliminated from our model because we consider a {\it test} particle
which, by definition, cannot modify the background spacetime.  Since
$\mathcal{H}^{(-)}$ and $\mathcal{H}^{(+)}$, being vector spaces,
must include the zero solutions, the Hilbert space
$\mathcal{H}^{(-)}\bigoplus\mathcal{H}^{(+)}$ cannot model the
quantum phase of our system.

\subsection{Regularization}

In the RCM space the quantum Hamiltonian, $\hat{H}_\epsilon$, for any
$t \in [-T,T]$, reads (we use the gauge $C_\epsilon =2$)
\begin{equation}
\hat{H_\epsilon}= \frac{\sqrt{t^2 + \epsilon^2 + \epsilon^2 r^2}}{t^2
+\epsilon^2}\; \frac{\partial^2}{\partial\theta^2} - \frac{t^2
+\epsilon^2}{\sqrt{t^2 + \epsilon^2 + \epsilon^2
r^2}}\;\frac{\partial^2}{\partial t^2} - \nonumber
\end{equation}
\begin{equation}\label{rh}
\frac{t(t^2 + \epsilon^2 + 2 \epsilon^2 r^2)}{(t^2 + \epsilon^2 +
\epsilon^2 r^2)^{3/2}}\;\frac{\partial}{\partial t} - \sqrt{t^2 +
\epsilon^2 + \epsilon^2 r^2}\; m^2
\end{equation}
Since the Hamiltonian is equal to the first-class constraint,  the
physical states are solutions to the equation
\begin{equation}\label{rh1}
\hat{H_\epsilon}\Psi=0 .
\end{equation}

\begin{figure}[h]
\centering \subfigure[]{

\includegraphics[width=2.3in]{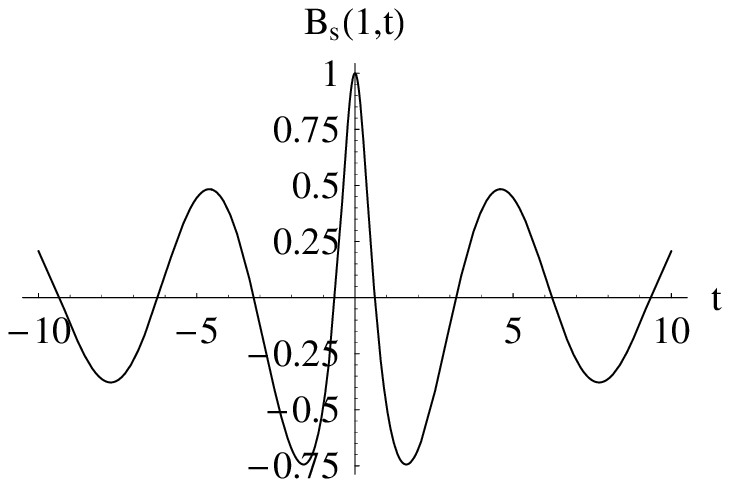}}
\hspace{0.4in} \subfigure[]{

\includegraphics[width=2.3in]{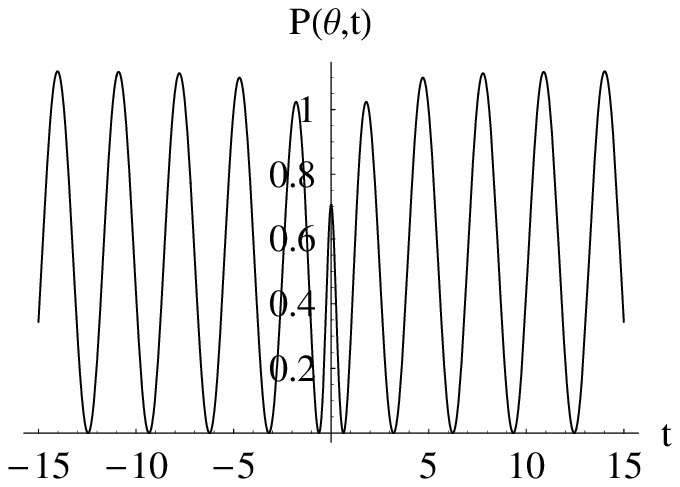}}
\caption{Properties of the symmetric solution with $\epsilon =0.5,\;
\rho=1$ and $r=1$: (a) the solution corresponding to $B_s (1,0)=1$
and $dB_s (1,0)/dt = 0$ as the initial values for (\ref{eqtr}), (b)
the probability density $P(\theta,t)$ corresponding to $A_1 (1,0)$
and $B_s (1,t)$.}

\end{figure}

\begin{figure}[h]
\centering \subfigure[]{

\includegraphics[width=2.3in]{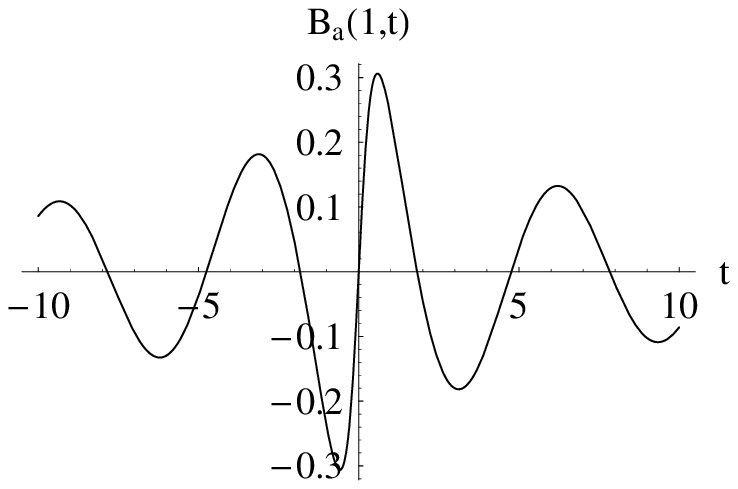}}
\hspace{0.4in} \subfigure[]{

\includegraphics[width=2.3in]{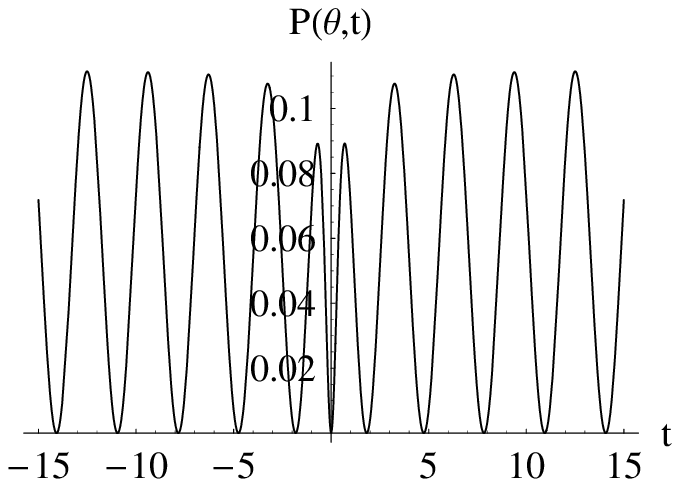}}
\caption{Properties of the anti-symmetric solution with $\epsilon
=0.5,\;\rho=1$ and $r=1$: (a) the solution corresponding to $B_a
(1,0)=0$ and $dB_a (1,0)/dt = 1$ as the initial values for
(\ref{eqtr}), (b) the probability density $P(\theta,t)$ corresponding
to $A_1 (1,0)$ and $B_a (1,t)$.}
\end{figure}

As in the case of $\epsilon =0$, the space of solutions to
(\ref{rh1}) defines the domain of boundedness of $\hat{H_\epsilon}$.
Substitution $\Psi(\theta,t)=A(\theta)B(t)$ into (\ref{rh1}) yields
\begin{equation}\label{eqthr}
\frac{d^2A}{d\theta^2}+\rho^2A=0,~~~~\rho\in \dR,
\end{equation}
and
\begin{equation}\label{eqtr}
\frac{(t^2+\epsilon^2)^2}{t^2+\epsilon^2+r^2\epsilon^2}\;\frac{d^2B}{dt^2}+
\frac{t(t^2+\epsilon^2)(t^2+\epsilon^2+2r^2\epsilon^2)}
{(t^2+\epsilon^2+2r^2\epsilon^2)^2}\;\frac{dB}{dt}+m^2(t^2+\epsilon^2+\rho^2)B=0.
\end{equation}

The equation (\ref{eqthr}) looks the same as in the non-regularized
case (\ref{eqth}) so the two independent solutions on $\dS^1$ read
\begin{equation}\label{solthr}
     A_1(\rho,\theta)= \pi^{-1/2} \cos(\rho\theta),~~~~~A_2(\rho,\theta)= \pi^{-1/2}
     \sin(\rho\theta),
\end{equation}
where  $r\rho = 0,\pm 1,\pm 2,\ldots$ (orthogonality and continuity
conditions).

Equation (\ref{eqtr}) is non-singular in $ [-T,T] $ so for each
$\rho$ it has two independent solutions which are bounded and smooth
in the entire interval. One may represent these solutions by a
symmetric and an anti-symmetric functions. We do not try to find the
analytic solutions to (\ref{eqtr}). What we really need to know are
general properties of them. In what follows  we further analyze only
numerical solutions to (\ref{eqtr}) by making use of \cite{SWM}.

Since in the regularized case the solutions are continuous in the
entire interval $[-T,T]$, the problem of gluing the solutions (the
main problem in  case $\epsilon = 0$) does not occur at all. Thus,
the construction of the Hilbert space, $\mathcal{H}_\epsilon$, by
using the space of solutions to (\ref{eqthr}) and (\ref{eqtr}) is
straightforward. The construction of the basis in
$\mathcal{H}_\epsilon$ may be done by analogy to the construction of
the basis in $\mathcal{H}^{(-)}$ (described in the subsection B). The
only difference is that now $t \in [-T,T]$ and instead of
(\ref{solt}) we use the solutions to (\ref{eqtr}). Let us denote them
by $B_i(\rho,mt)$, where $i=s$ and $i=a$ stand for symmetric and
anti-symmetric solutions, respectively. Figures 6 and 7 present two
examples of solutions to (\ref{eqtr}) for $\rho =1$ and the
corresponding probability densities. We can see that $P(\theta,t)$ is
a bounded and continuous function on $[-T,T]\times \dS^1$, as in the
case of $\epsilon =0$ (cp with Figs. 3 and 4).

We define the scalar product as follows
\begin{equation}\label{sc}
    <\psi_1|\psi_2> := \int_\Gamma d \mu \;\overline{\psi}_1 \;\psi_2,~~~~~~d\mu
    :=\sqrt{-g}\; d\theta \;dt = \sqrt{t^2+\epsilon^2+r^2\epsilon^2}\; d\theta \;dt,
\end{equation}
where ${\Gamma}:= [-T,T] \times \dS^1$, and where an explicit form of
$d\mu$ is found by making use of (\ref{line3}).

It is evident that $\hat{H}_\epsilon$ is self-adjoint on
$\mathcal{H}_\epsilon$. The main difference between the deterministic
case with $\epsilon =0$ and the present case $\epsilon >0$ is that in
the former case the Hilbert space is one dimensional ($\rho =0$),
whereas in the latter case it is $2\times 2 \times \infty$
dimensional ($r\rho \in \dZ$).

\section{Summary and conclusions}

The Cauchy problem at the cosmological singularity of the geodesic
equations may be `resolved' by the regularization, which replaces
the double conical vertex of the CM space by a space with the
vertex of the big-bounce type, i.e. with non-vanishing space
dimension at the singularity. We have presented a specific example
of such regularization of the CM space. Both classical and quantum
dynamics of a particle in the  regularized CM space  are
deterministic and stable. We have examined these aspects of the
dynamics at the phase space and Hamiltonian levels. The classical
and quantum dynamics of a particle in the regularized CM space is
similar to the dynamics in the de Sitter space
\cite{WP,Piechocki:2003hh}. We are conscious that our
regularization of the singularity is rather {\it ad hoc}. Our
arguing (presented at the beginning of Sec. IIB) that taking into
account the interaction of a  physical particle with the
singularity may lead effectively to changing of the latter into a
big-bounce type singularity should be replaced by analyzes.
However, examination of this problem is beyond the scope of the
present paper, but will be considered elsewhere.

The classical dynamics in the CM space is unstable (apart from the
one class of geodesics). However, the quantum dynamics is well
defined. The Cauchy problem of the geodesics is not an obstacle to
the quantization. The examination of the quantum stability has
revealed surprising result that in one case a quantum particle
propagates deterministically in the sense that it can be described by
a quantum state  that is continuous at the singularity. This case is
very interesting as it says that there can exist deterministic link
between the data of the pre-singularity and post-singularity epochs.
All other states have discontinuity at the singularity of the CM
space, but they can be used successfully to construct a Hilbert
space. This way we have proved the stability of the dynamics of a
{\it quantum} particle.

At the quantum level the stability condition requiring the
boundedness from below of the Hamiltonian operator means  the
imposition of the first-class constraint onto the space of quantum
states to get the space of  {\it physical}  quantum states. The
resulting equation depends on all spacetime coordinates. In the
pre-singularity and post-singularity epochs the CM space is locally
isometric to the Minkowski space \cite{Malkiewicz:2005ii}. Owing to
this isometry, the stability condition is in fact  the Klein-Gordon,
KG, equation. The space of solutions to the KG equation in these two
epochs and the corresponding Hilbert space are fortunately
non-trivial ones, otherwise our quantum theory of a particle would be
 empty.

Quantization of the phase space carried out in Sec. IV (and in our
previous paper \cite{Malkiewicz:2005ii}), corresponds to some
extent to the method of quantization in which one first solves
constraints at the classical level and then quantize the resulting
theory. Quantization that we call here examination of the
stability at the quantum level, is effectively the method in which
we impose the constraint, but at the quantum level. The results we
have obtained within both methods of quantization are consistent.
It means that the quantum theory of a particle in the compactified
Milne space does exist. The CM space seems to model the
cosmological singularity in a satisfactory way\footnote{Our result
should be further confirmed by the examination of the dynamics of
a particle in a higher dimensional CM space.}.

It turns out that the  time-like geodesics of our  CM space may
have interpretation in terms of cosmological solutions of some
sophisticated higher dimensional field theories
\cite{Russo:2004am,Bergshoeff:2005cp,Bergshoeff:2005bt}. We have
already discussed some aspects of this connection in our previous
paper (see Sec. 5 of \cite{Malkiewicz:2005ii}). Presently, we can
say that in one case (see Sec. 4 of \cite{Russo:2004am} and Sec.
V.B.1 of this paper) this analogy extends to the quantum level:
transition of a particle through the cosmological singularity in
both models is mathematically well defined. In both cases the
operator constraint is used to select quantum physical states.
Elaboration of this analogy needs an extension of our results to
the Misner space (that consists of the Milne and Rindler spaces)
because it is the spacetime  used in \cite{Russo:2004am}. Another
subtlety is connected with the fact that we carry out analysis in
the compactified space, whereas the authors of \cite{Russo:2004am}
use the covering space.

There exists another model to describe the  evolution of the
universe based on string/M theory. It is called the pre-big-bang
model \cite{Gasperini:2002bn}. However, the ST model is more
self-consistent and complete.

Other sophisticated model called loop quantum cosmology, LQC, is
based on non-perturbative formulation of quantum gravity called
the loop quantum gravity, LQG \cite{TT,CR}. It is claimed that the
CS is resolved in this approach \cite{Bojowald:2006da}. However,
this issue seems to be still open due to  the assumptions made in
the process of truncating the infinite number degrees of freedom
of the LQG to the finite number used in the LQC
\cite{Brunnemann:2005in,Brunnemann:2005ip}. This model has also
problems in obtaining an unique semi-classical approximations
\cite{Nicolai:2005mc}, which are required to link the quantum
phase with the nearby classical phase in the evolution of the
Universe. For response to \cite{Nicolai:2005mc} we recommend
\cite{Thiemann:2006cf}.

Quantization of dynamics of {\it extended} objects in the CM space
is our  next step. There exist  promising results on propagation
of a string and membrane \cite{Turok:2004gb,Niz:2006ef}. However,
these results concern extended objects in the low energy states
called the zero-mode states and quantum evolution is approximated
by a semi-classical model. Recently, we have quantized the
dynamics of a string in the  CM space rigorously
\cite{Malkiewicz:2006bw}, but our results concern only the
zero-mode state of a string.  For drawing firm conclusions about
the physics of the problem, one should also examine the non-zero
modes. Work is in progress.

\begin{acknowledgments}
We would like to thank L. {\L}ukaszuk and N. Turok for valuable
discussions, and the anonymous referees for constructive
criticisms.
\end{acknowledgments}

\end{document}